\documentclass[12pt]{iopart}

\usepackage{amstext}
\usepackage{epsfig}
\usepackage{graphicx}

\begin{document}

\title[]{Phenomenology of the three-flavor PNJL model
           and thermal strange quark production}

\author{Hung-Ming Tsai and Berndt M\"uller}

\address{Department of Physics, Duke University, Durham, NC 27708, USA}
\ead{ht25@phy.duke.edu, mueller@phy.duke.edu}

\begin{abstract}
We study the temperature dependence of the adjoint Polyakov loop and
its implication for the momentum spectrum of gluons in the
mean-field approximation. This allows us to calculate the
contribution of the thermal (transverse) gluons to the thermodynamic
pressure. As an application, we evaluate the rates for the strange
quark pair-production processes $q\bar{q} \to s\bar{s}$ and $gg \to
s\bar{s}$ as functions of temperature including thermal effects on
quark deconfinement and chiral symmetry breaking.
\end{abstract}


\maketitle

\section{Introduction}

The increased abundance of particles containing strange quarks in
the spectrum of emitted hadrons, especially hyperons, was proposed
by Hagedorn and Rafelski \cite{Hagedorn:1980kb} as a signal for the
formation of quark matter in relativistic heavy ion collisions. Soon
afterwards, the enhanced pair production of strange quarks required
for the saturation of strange quark phase space was predicted to
occur as an effect of quark and gluon deconfinement
\cite{Rafelski:1982pu,Biro:1982ud,Koch:1986ud,Matsui:1985eu}. The
predicted enhancement has been observed in many experiments (see, in
particular: \cite{Andersen:1999ym,Abelev:2007xp}). The parameter
$\gamma _s$, describing the degree of saturation of the phase space
of strange hadrons, has been determined by thermal chemical fits to
the abundances of hadrons emitted in collisions between two
$^{197}$Au nuclei at center-of-mass energies of 200 GeV/nucleon at
the Relativistic Heavy Ion Collider (RHIC). Values for this
parameter obtained in such fits range from $\gamma _s = 1.03\pm
0.04$ \cite{Kaneta:2004zr} to $\gamma _s = 2.00\pm 0.02$
\cite{Letessier:2005qe}.

It is important to understand how the dynamics of deconfinement and
chiral symmetry affects the prediction of production of the strange
quarks and therefore its enhancement. The recently developed
three-flavor PNJL model \cite{Ciminale:2007sr,Fukushima:2008wg} has
made such a study possible. In the present work we explore the
effect of the temperature dependence of the Polyakov loops and
chiral condensates on the strange quark production. Our article is
structured as follows. After stating the basic equations of the PNJL
model, including the thermal quark and antiquark distribution
functions, we obtain an explicit fit for the temperature dependence
of the effective action of the Polyakov loop in mean-field theory.
We confirm that this action satisfies the scaling of the thermal
averages of the Polyakov loop in different color representations by
their Casimir operator. We then calculate the temperature dependence
of the thermal average of the adjoint Polyakov loop, the thermal
distribution function of transverse gluons and the contribution of
transverse gluons to the thermodynamic potential.

Finally, we calculate the temperature dependence of the
pair-production rate of strange quarks using the Polyakov
loop-suppressed quark and gluon distribution functions. With the aid
of the PNJL model, we identify the temperature where the gluonic
contribution to the production rate becomes dominant. The cross-over
of the contributions from light quarks and gluons is a novel
phenomenon which does not exist in the traditional approach to
strange quark-pair production based on perturbation theory
\cite{Rafelski:1982pu}.

\section{The PNJL model}

The phase transformations of QCD matter due to deconfinement and
chiral symmetry restoration have been combined in one theoretical
framework, which is the Nambu-Jona-Lasinio model with the Polyakov
loop (PNJL model) \cite{Fukushima:2003fw,Ratti:2005jh}. The Polyakov
loop in color-SU(3) representation $r$ is defined as
\begin{eqnarray}
L_r = \mathcal{P} \exp \left( {ig\int_0^{1/T} {d\tau A_4 \left(
{{\mathrm {\bf x}},\tau } \right)} } \right),\label{eq:P-loop-def}
\end{eqnarray}
\noindent where $\mathcal{P}$ denotes that the exponential is
path-ordered, $T$ denotes the temperature and $A_4 \left( {{\mathrm
{\bf x}},\tau } \right)$ is the temporal component of the
$\mathrm{SU}(3)$ gauge field in representation $r$. In particular,
$L_3 $ and $L_8 $ denote the Polyakov loops in the fundamental and
adjoint representations, respectively. The traces of the Polyakov
loops are defined as
\begin{eqnarray}
\ell _3 = N_c^{-1} \tr_F L_3 , \; \; \; \; \; \bar {\ell }_3 =
N_c^{-1} \tr_F L_3^\dag ,  \label{eq:ell_3-def}\\
\ell _8 = (N_c^2 - 1)^{-1} \tr_A L_8, \label{eq:ell_8-def}
\end{eqnarray}
where $\tr_F $ and $\tr_A $ denote the color traces in the
fundamental and adjoint representation, respectively. Note that the
dependence of $L_r$, $L_r^\dag$, $\ell_r$ and $\bar{\ell}_r$ on the
spatial coordinate $\mathbf{x}$ is suppressed in
(\ref{eq:P-loop-def})-(\ref{eq:ell_8-def}). With the above
definitions, the Lagrangian of the three flavor PNJL model is
\cite{Fukushima:2008wg} \noindent
\begin{eqnarray}
\mathcal{L} &=& \bar {\psi }\left( {i\gamma \cdot D - \hat {m}_0 }
\right)\psi - \mathcal{U} ( \ell_3, \bar{\ell}_3 ; T )
 \nonumber \\
&& + \frac{g_S }{2} \sum_{a=0}^{8} \left[ {\left( {\bar {\psi
}\lambda ^a\psi } \right)^2 + \left( {\bar {\psi }i\gamma _5 \lambda
^a\psi } \right)^2} \right] \nonumber \\
&& + g_D \left[ {\det \bar {\psi }\left( {1 - \gamma _5 }
\right)\psi + h.c.} \right], \label{eq:PNJL-Lagr}
\end{eqnarray}
\noindent where $D_\mu = \partial _\mu - g\delta _{\mu 4} A_4$ is
the gauge-covariant derivative, $\mathcal{U} ( \ell_3, \bar{\ell}_3
; T )$ is the effective potential for the Polyakov loop and the
three-flavor current quark mass matrix $\hat {m}_0 =
\mathrm{diag}(m_{u,0} ,m_{d,0} ,m_{s,0} )$. In the limit of isospin
symmetry, $m_{u,0} = m_{d,0} = m_{q,0} $. In the mean-field
approximation, the chiral condensates and the thermal averages of
the Polyakov loops are the order parameters of the phase transition
\cite{Fukushima:2008wg,Ratti:2005jh}. Note that, in the mean-field
approximation, the thermal average of the Polyakov loop is
independent of the spatial coordinate $\mathbf{x}$.

To encode the features of the temperature dependence of the
effective potential $\mathcal{U} ( \ell_3, \bar{\ell}_3 ; T )$, the
nonperturbative contribution to the gluon thermodynamic potential
per unit volume is assumed to be of the following form
\cite{Fukushima:2008wg}:
\begin{eqnarray}
\Omega_{g}^{\mathrm{NP}} &=& - b \, T\left\{ 54 \,
\mathrm{exp}\left( - \frac{a}{T} \right) \langle \ell _3 \rangle
\langle \bar {\ell }_3
\rangle  \right. \nonumber \\
&&+\left. \ln \left[ {1 - 6 \langle \ell _3 \rangle \langle \bar
{\ell }_3 \rangle - 3\left( { \langle \ell _3 \rangle \langle \bar
{\ell }_3 \rangle } \right)^2 + 4\left( { \langle \ell _3 \rangle ^3
+ \langle \bar {\ell }_3 \rangle ^3 } \right)} \right] \right\},
\label{eq:Omega_g_NP}
\end{eqnarray}
\noindent where $\langle {\ell _3 } \rangle $ and $\langle {\bar
{\ell}_3 } \rangle $ are the thermal averages of the Polyakov loops
in (\ref{eq:ell_3-def}). Standard values for the parameters in
(\ref{eq:PNJL-Lagr}) and (\ref{eq:Omega_g_NP}) are $a = 0.664$~GeV,
$b = 0.03 \Lambda^3$, $g_S = 3.67 \Lambda ^{-2}$, $g_D = - 9.29
\Lambda ^{-5}$ and $\Lambda = 0.6314$~GeV \cite{Fukushima:2008wg}.
Moreover, it is straightforward to obtain the quark partition
function from the quark Lagrangian, yielding the quark grand
canonical thermodynamic potential per unit volume
\cite{Fukushima:2008wg}:
\begin{eqnarray}
\Omega _{q} &=& - 2T \left\langle \sum\limits_{f = u,d,s} \int
\frac{d^3k}{(2\pi )^3}  \left\{  \tr_F   \ln \left[ 1 + L_3
\exp\left( {{ - (E_f(\mathbf{k}) -\mu _f ) \over T}} \right) \right]
\right. \right.
\nonumber\\
&&  \left. \left.+   \tr_F \ln \left[ 1 + L_3^{\dagger} \exp\left(
{{ - ( E_f(\mathbf{k}) +\mu _f ) \over T}} \right) \right] \right\}
\right\rangle
\nonumber\\
&& -6\sum\limits_{f = u,d,s} \int \frac{d^3k}{(2\pi)^3}
E_f(\mathbf{k})  \theta \left(\Lambda ^2 - |\mathrm {\bf k}|^2
\right), \label{eq:grand-q}
\end{eqnarray}
where $E_f(\mathbf{k})  = ({|\mathrm {\bf k}|}^2 + m_f^2 )^{1/2}$
are the single-particle energies, and $m_f $ and $\mu _f $ are the
constituent mass and chemical potential of the quark with flavor
$f$. The constituent quark masses are related to the current masses
and the chiral condensates by
\begin{eqnarray}
m_q &=& m_{q,0} - 2g_S \langle {\bar {q}q} \rangle - 2g_D \langle
{\bar {q}q} \rangle \langle {\bar {s}s} \rangle, \qquad (q=u,d)
\label{eq:mq}
\\
m_s &=& m_{s,0} - 2g_S \langle {\bar {s}s} \rangle - 2g_D \langle
{\bar {q}q} \rangle ^2,\label{eq:ms}
\end{eqnarray}
where  $\langle {\bar {u}u} \rangle = \langle {\bar {d}d} \rangle $
and $m_{u,0} = m_{d,0}$. Furthermore, in the mean-field
approximation, the quark condensate contribution to the
thermodynamic potential per unit volume is \cite{Fukushima:2008wg}
\begin{eqnarray}
\Omega_{\mathrm{cond}} = g_S \left( 2 \langle \bar{q} q \rangle^2 +
\langle \bar{s} s \rangle^2 \right)+4 g_D \langle \bar{q} q
\rangle^2 \langle \bar{s} s \rangle. \label{eq:Omega_cond}
\end{eqnarray}

\noindent In the present article, the parameters are chosen as
follows: $m_{q,0} = 0.0055$~GeV, $m_{s,0} = 0.1357$~GeV. The vacuum
quark condensates are $\langle {\bar {q}q} \rangle _0 = \left( { -
0.246\,\;\mathrm{GeV}} \right)^ 3$ and $\langle {\bar {s}s} \rangle
_0 =\left( { - 0.267\,\;\mathrm{GeV}} \right)^ 3$
\cite{Fukushima:2008wg}.

In the mean-field approximation, one easily derives the temperature
dependence of the chiral condensates $\langle\bar{q}q\rangle$ and
$\langle\bar{s}s\rangle$ as well as the thermal averages of the
Polyakov loops in the fundamental representation, i.e. $\langle
{\ell _3 } \rangle $ and $\langle {\bar {\ell}_3 } \rangle $
\cite{Fukushima:2008wg}. Furthermore, it is straightforward to
derive from (\ref{eq:grand-q}) the color-averaged distribution
functions for $q$ and $\bar {q}$ \cite{Hansen:2006ee}:
\begin{eqnarray}
f_{q} \left( k \right) &=& \frac{ \langle {\ell _3 } \rangle
\lambda_{+} + 2 \langle {\bar \ell _3 } \rangle \lambda_{+}^2 +
\lambda_{+}^3 }{1 + 3 \langle {\ell _3 } \rangle \lambda_{+} + 3
\langle {\bar \ell _3 } \rangle \lambda_{+}^2 + \lambda_{+}^3},
\label{eq:quark-dist}
\\
f_{\bar {q}} \left( k \right) &=& \frac{\langle {\bar \ell _3 }
\rangle \lambda_{-} + 2 \langle {\ell _3 } \rangle \lambda_{-}^2 +
\lambda_{-}^3 }{1 + 3\langle {\bar \ell _3 } \rangle \lambda_{-} +
3\langle {\ell _3 } \rangle \lambda_{-}^2 + \lambda_{-}^3},
\label{eq:antiquark-dist}
\end{eqnarray}
where we introduced the abbreviations
$\lambda_{\pm}=\mathrm{exp}[-((|\mathrm {{\bf k}}|^2 + m_q^2 )^{1 /
2} \mp \mu)/T]$ and chose an isospin-independent chemical potential
for the light quarks: $\mu = \mu _u = \mu _d$.
Equations~(\ref{eq:quark-dist}) and (\ref{eq:antiquark-dist}) are
obtained by assuming the Weiss mean-field approximation
\cite{Fukushima:2008wg}. In \ref{sec:appendix}, we discuss several
Polyakov-loop averaging procedures for evaluating the quark
thermodynamic potential and thereby comparing the quark distribution
functions derived from these different averaging procedures. In the
following, we will consider the choice $\mu = 0.1$~GeV. These
distribution functions are to be used in the study of the strange
quark pair-production rates due to the process $q\bar {q} \to s\bar
{s}$.

\section{Adjoint Polyakov loop and gluon distribution function}

In order to describe the process $gg \to s\bar{s}$, we also need to
calculate the adjoint Polyakov loop and study how it affects the
gluon distribution function. Starting from the Yang-Mills Lagrangian
${\mathcal L}_{g} = - (1 / 4)F_{\mu \nu }^a F^{a,\mu \nu}$, it
straightforward to write down the gluon partition function for the
transverse gluons, and obtaining the perturbative contribution to
the gluon thermodynamic potential per unit volume
\cite{Meisinger:2001cq,Megias:2004hj},
\begin{eqnarray}
\Omega _{g}^{\mathrm{P}} &=& 2T \left\langle \int \frac{d^3k}{(2\pi )^3}
\tr_A \ln \left[ 1 - L_8\, \exp( - \left| {\mathrm {\bf k}} \right| / T) \right] \right\rangle
\nonumber \\
&&+ 8\int {\frac{d^3k}{(2\pi )^3}\left| {\mathrm {\bf k}}
\right|\theta \left( {\Lambda ^2 - | \mathrm {\bf k}} |^2 \right)}.
\label{eq:Omega_g_P1}
\end{eqnarray}
Because $A_4({\bf x},\tau)$ is assumed to be independent of the
spatial coordinate ${\mathrm {\bf x}}$  in the mean-field
approximation, the Polyakov loop in the fundamental representation
can be gauge rotated to diagonal form, $L_3 = \mathrm{diag}(e^{i\phi
_1 },e^{i\phi _2 },e^{i\phi _3 })$, with $\phi _3 = - (\phi _1 +
\phi _2 )$. Therefore,
\begin{eqnarray}
\ell_3 = \left( \bar{\ell}_3 \right)^{\ast}=
\frac{1}{3}[\exp(i\phi_1)+\exp(i\phi_2)+\exp(-i(\phi_1+\phi_2))].
\label{eq:ell_3-MFA}
\end{eqnarray}

\noindent In the same gauge, $L_8 $ can be expressed in terms of the
eigenvalues of $L_3 $ \cite{Meisinger:2001cq}:
\begin{eqnarray}
L_8 &=& \mathrm{diag}\left(1,1,e^{ i\phi _{31}}, e^{i\phi _{13}},
e^{ i\phi _{23}}, e^{i\phi _{32}}, e^{ i\phi _{21}}, e^{i\phi _{12}} \right),
\label{eq:L_8}
\end{eqnarray}
where $\phi_{jk}=\phi_j-\phi_k$. Thus,
\begin{equation}
\ell_8 = \frac{1}{8}\tr_A L_8
= \frac{1}{4}\left( 1+ \sum\limits_{j< k} \cos \phi _{jk} \right) .
\label{eq:ell_8}
\end{equation}
Inserting (\ref{eq:L_8}) into (\ref{eq:Omega_g_P1}) and after some
algebraic transformations, the color-averaged gluon distribution
function is obtained as
\begin{equation}
f_g \left( k \right) = \frac{1}{8}\sum\limits_{n = 1}^\infty
{\langle {\tr_A L_8^n } \rangle \exp \left( { - {n\left|
{\mathrm {\bf k}} \right| / T}} \right)} ,
\label{eq:gluon-dist}
\end{equation}
where
\begin{equation}
\tr_A L_8^n = 2\left[ 1 + \sum\limits_{j < k} \cos n\phi _{jk}
\right].
\end{equation}

\begin{figure}[tb]
\centerline{\epsfig{file=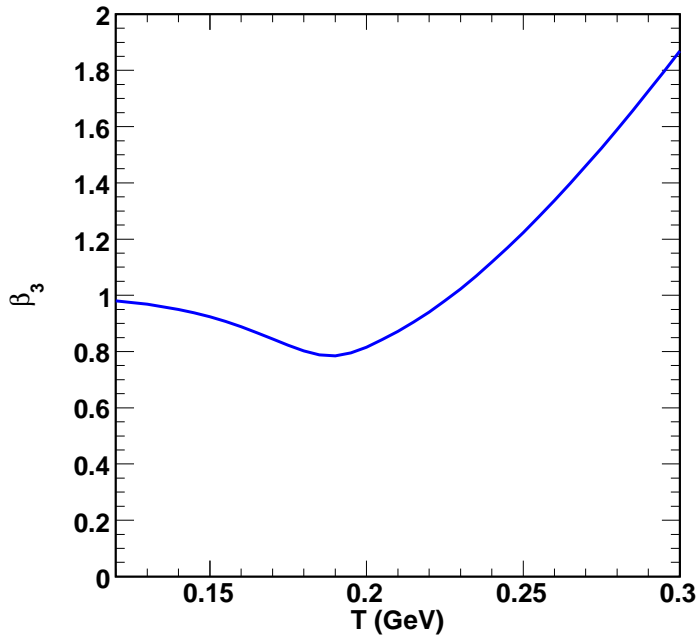,width=10cm}}
\caption{The fit parameter $\beta_3$ as a function of temperature.}
\label{fig:beta3}
\end{figure}

The remaining task is to evaluate $\langle {\tr_A L_8^n } \rangle $.
In order to do so, we need to specify the full distribution of
eigenvalues of the Polyakov loop, i.~e.\ the distribution of phases
$\phi_i$. The thermal average of any function of the eigenvalues of
the Polyakov loop, $F(\phi _1 ,\phi _2 )$, is defined as
\begin{eqnarray}
\langle {F(\phi _1 ,\phi _2 )} \rangle = \frac{\int_0^{2\pi }
{d\phi _1 \int_0^{2\pi } {d\phi _2 H(\phi _1 ,\phi _2)
W(\phi _1 ,\phi _2 ;T) F(\phi _1 ,\phi_2)} } }
{\int_0^{2\pi } {d\phi _1 \int_0^{2\pi } {d\phi _2
H(\phi _1 ,\phi _2) W(\phi _1 ,\phi _2 ;T)} } },
\label{eq:expectation}
\end{eqnarray}
where $H(\phi _1 ,\phi _2 )$ is the SU(3) Haar measure, the weight
function $W(\phi _1 ,\phi _2 ;T)$ denotes the distribution of
eigenvalues of the Polyakov loop. The Haar measure for the SU(3)
symmetry group is given by $H(\phi_1 ,\phi _2 ) = \prod\nolimits_{j
< k} \sin ^2(\phi _{jk}/ 2)$. Since $\phi _3 = - (\phi _1 + \phi _2
)$, the integration in (\ref{eq:expectation}) only goes over $\phi
_1 $ and $\phi _2 $. A suitable choice of the weight function is a
crucial step in evaluating $\langle {F(\phi _1 ,\phi _2 )}\rangle $.
We follow Gocksch and Ogilvie \cite{ Gocksch:1984yk} and Gupta {\em
et al.} \cite{Gupta:2007ax} in choosing a weight function of the
form
\begin{eqnarray}
W(\phi _1 ,\phi _2 ;T) = \exp \left( {6 \, d \, \beta
_3 \, \langle {\ell _3 } \rangle \, \mathrm{Re}(\ell _3)} \right),
\label{eq:weight}
\end{eqnarray}
where $d = 3$ and $\beta_3 ( T )$ is a fit parameter depending on
temperature. The particular form (\ref{eq:weight}) is suggested by
the strong coupling expansion of the gauge theory. We note that
$\exp[-\mathcal{U}(\ell_3,\bar{\ell}_3;T)/(b \, T)]$ and $H(\phi_1,
\phi_2)W(\phi_1, \phi_2)$ have a corresponding structure expressed
in terms of the eigenvalues of the fundamental Polyakov loops, as
can be seen from assuming $\mathcal{U}(\ell_3,\bar{\ell}_3;T)$ to be
in the form of (5) with $\langle \ell_3 \rangle$ and $\langle
\bar{\ell}_3 \rangle$ replaced by $\ell_3$ and $\bar{\ell}_3$
respectively. Starting from (\ref{eq:weight}) and inserting the
known values of $\langle {\ell _3 } \rangle $ into
(\ref{eq:expectation}), the temperature dependence of $\beta _3$ can
be solved numerically, as shown in figure~\ref{fig:beta3}. We now
have obtained an explicit expression for $W(\phi _1,\phi _2)$ at
each temperature.

We note that the temperature corresponding to the minimum of
$\beta_3$ in figure~\ref{fig:beta3} coincides with the critical
temperature of the deconfinement phase transition in the mean-field
approximation. In their investigation of the gluonic contribution to
the thermodynamic potential of the PNJL model, Megias {\em et al.}
\cite{Megias:2004hj} used a weight function of similar form as
(\ref{eq:weight}), but did not make the mean-field approximation.
They determined the parameter $\beta_3$ from the empirical relation
between string tension and the deconfinement temperature of the pure
gauge theory. Here, we have determined $\beta_3$ by imposing a
self-consistency condition on the expectation value of the
fundamental Polyakov loop.

\begin{figure}[tb]
\centerline{\epsfig{file=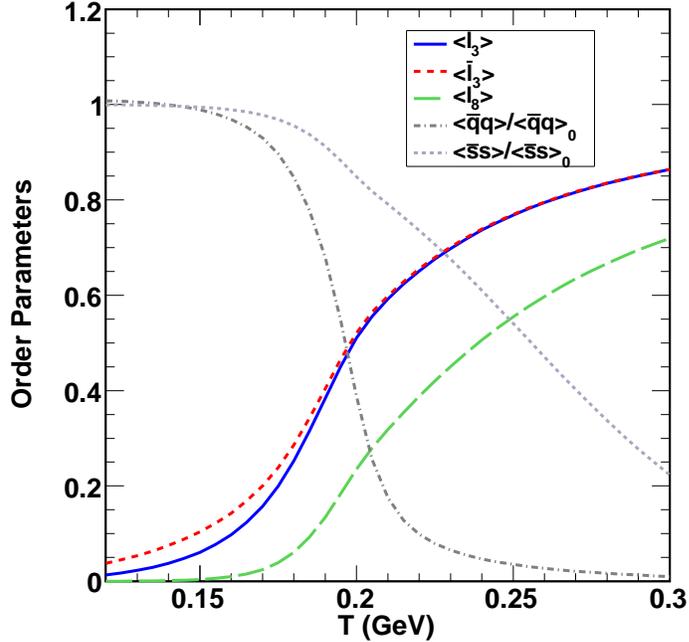,width=10cm}} \caption{The
temperature dependence of the order parameters, $\ell_3$,
$\bar{\ell}_3$, $\ell_8$, $\langle \bar{q}q \rangle/\langle \bar{q}q
\rangle_0$ and $\langle \bar{s}s \rangle/\langle \bar{s}s
\rangle_0$. The values of the vacuum quark condensates are $\langle
{\bar {q}q}\rangle _0 = \left( { - 0.246 \;\mathrm{GeV}} \right)^ 3$
and $\langle {\bar {s}s} \rangle _0 =\left( { - 0.267
\;\mathrm{GeV}} \right)^ 3$.} \label{fig:order_parameters}
\end{figure}

\begin{figure}[tb]
\centerline{\epsfig{file=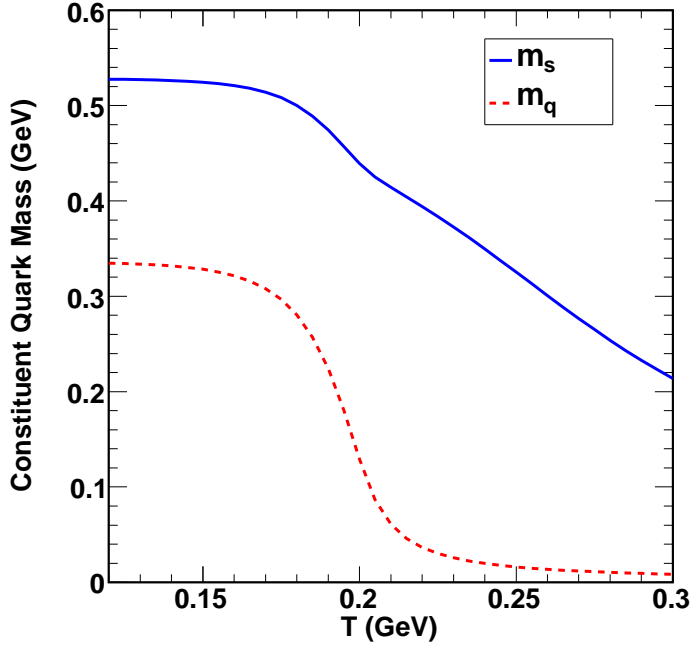,width=10cm}} \caption{The
temperature dependence of the constituent quark masses, $m_q$ and
$m_s$, in (\ref{eq:mq}) and (\ref{eq:ms}) respectively.}
\label{fig:masses}
\end{figure}

With the weight function (\ref{eq:weight}),
equation~(\ref{eq:expectation}) allows us to evaluate the thermal
average of the adjoint Polyakov loop $\langle {\ell _8 } \rangle $
and thus the gluon distribution function $f_g(k)$. The temperature
dependence of the quark condensates and the thermal averages of the
Polyakov loop $\langle {\ell _3 } \rangle $, $\langle {\bar \ell _3
}\rangle $ were first calculated in \cite{Fukushima:2008wg}, as well
as the temperature dependence of $\langle {\ell _8 } \rangle $ are
shown in figure~\ref{fig:order_parameters}. Furthermore, the
temperature dependence of the constituent masses of quarks are
obtained from (\ref{eq:mq}) and (\ref{eq:ms}), as shown in
figure~\ref{fig:masses}.

\begin{figure}[tb]
\centerline{\epsfig{file=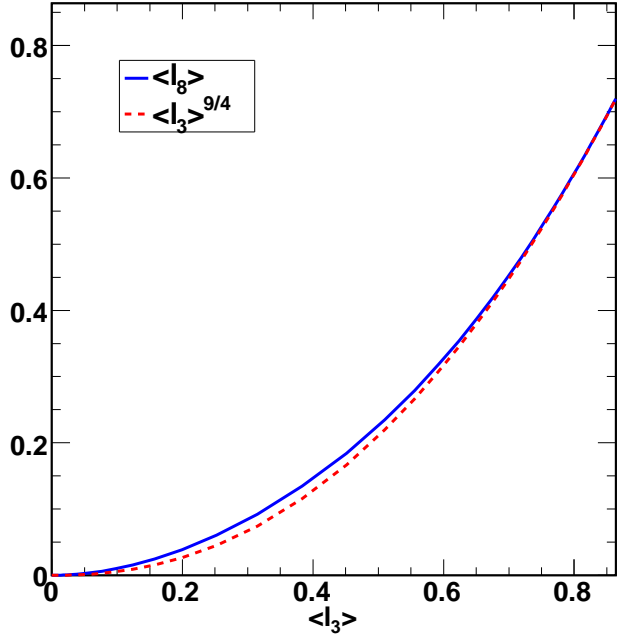,width=10cm}} \caption{The
relations of $\langle \ell_8 \rangle$ to $\langle \ell_3 \rangle$
and $\langle \ell_3 \rangle^{9/4}$ to $\langle \ell_3 \rangle$. The
temperature $T$ ranges from $0.12$~GeV to $0.30$~GeV. These two
curves indicates that the values of $\langle \ell_8 \rangle$ and
$\langle \ell_3 \rangle$ is consistent with Casimir scaling,
$\langle \ell_8 \rangle=\langle \ell_3 \rangle^{9/4}$.}
\label{fig:Casimir_scaling}
\end{figure}

To verify the validity of $\langle {\ell _8 } \rangle $ obtained by
this procedure, we check its consistency with the Casimir scaling of
the thermal averages of the Polyakov loop observed in lattice QCD
\cite{Gupta:2007ax,Gupta:2006qm}. Casimir scaling refers to a
relation, valid for all temperatures, between the thermal averages
of the Polyakov loop in different representations $r$ of color-SU(3)
of the form
\begin{equation}
\langle {\ell _r } \rangle = \langle {\ell _3 } \rangle ^{d_r },
\label{eq:Casimir}
\end{equation}
where $d_r = C_2 (r) / C_2 (3)$ and $C_2(r)$ denotes the eigenvalue
of the quadratic Casimir operator in representation $r$. For the
adjoint representation, $d_8 = C_2(8)/C_2(3) = 9 / 4$ and thus
$\langle {\ell _8 } \rangle = \langle {\ell _3 } \rangle ^{9/4}$.
Figure~\ref{fig:Casimir_scaling} shows our results for $\langle
{\ell _8 } \rangle$, together with $\langle {\ell _3 } \rangle ^{9 /
4}$, as function of $\langle {\ell _3 } \rangle $. As can be seen,
the values of $\langle {\ell _8 } \rangle $ and $\langle {\ell _3
}\rangle $ show a good agreement with Casimir scaling
(\ref{eq:Casimir}) for temperatures $T$ ranging from $0.12$~GeV to
$0.30$~GeV.

\begin{figure}[tb]
\centerline{\epsfig{file=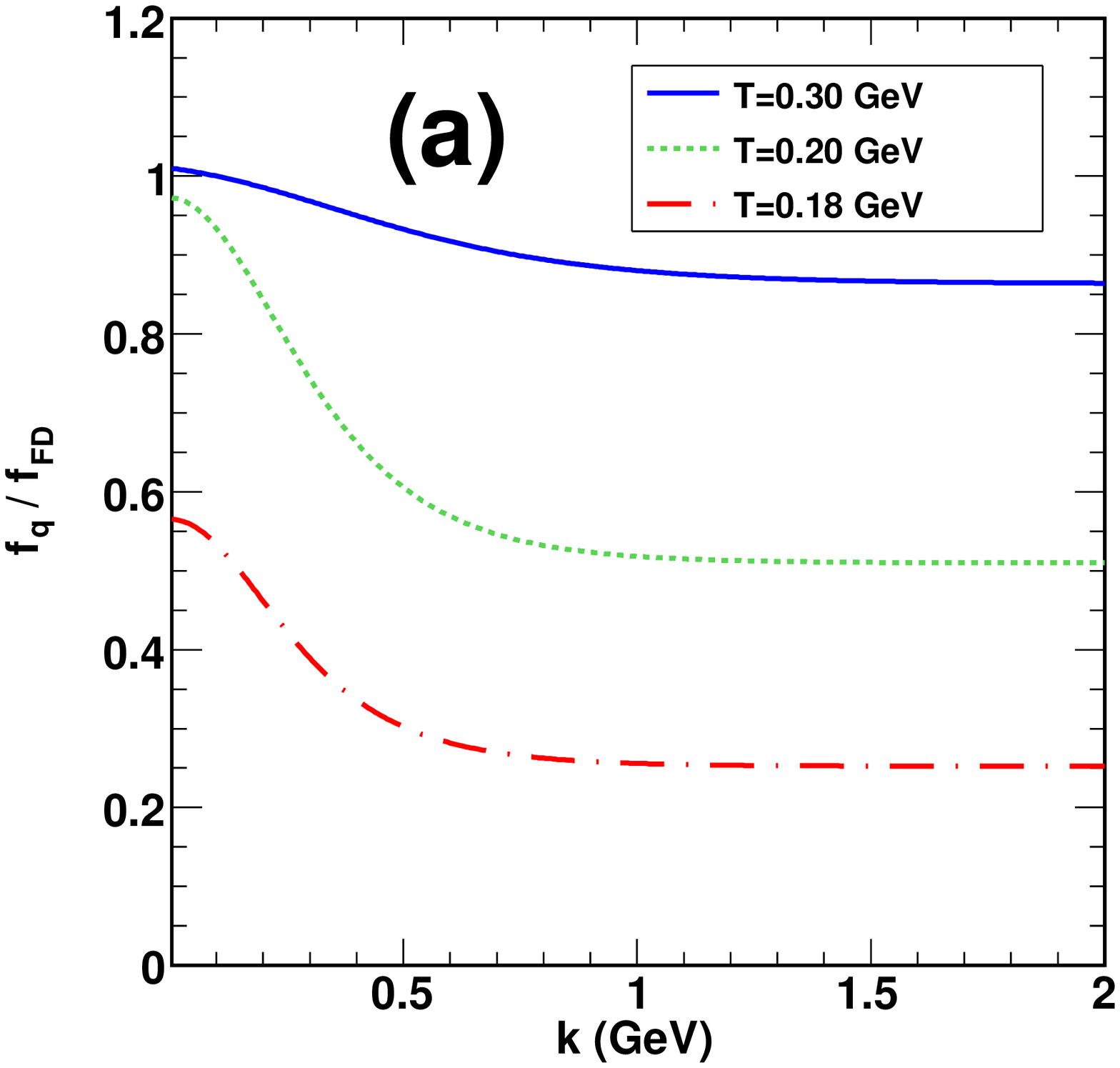,width=7.5cm}
\epsfig{file=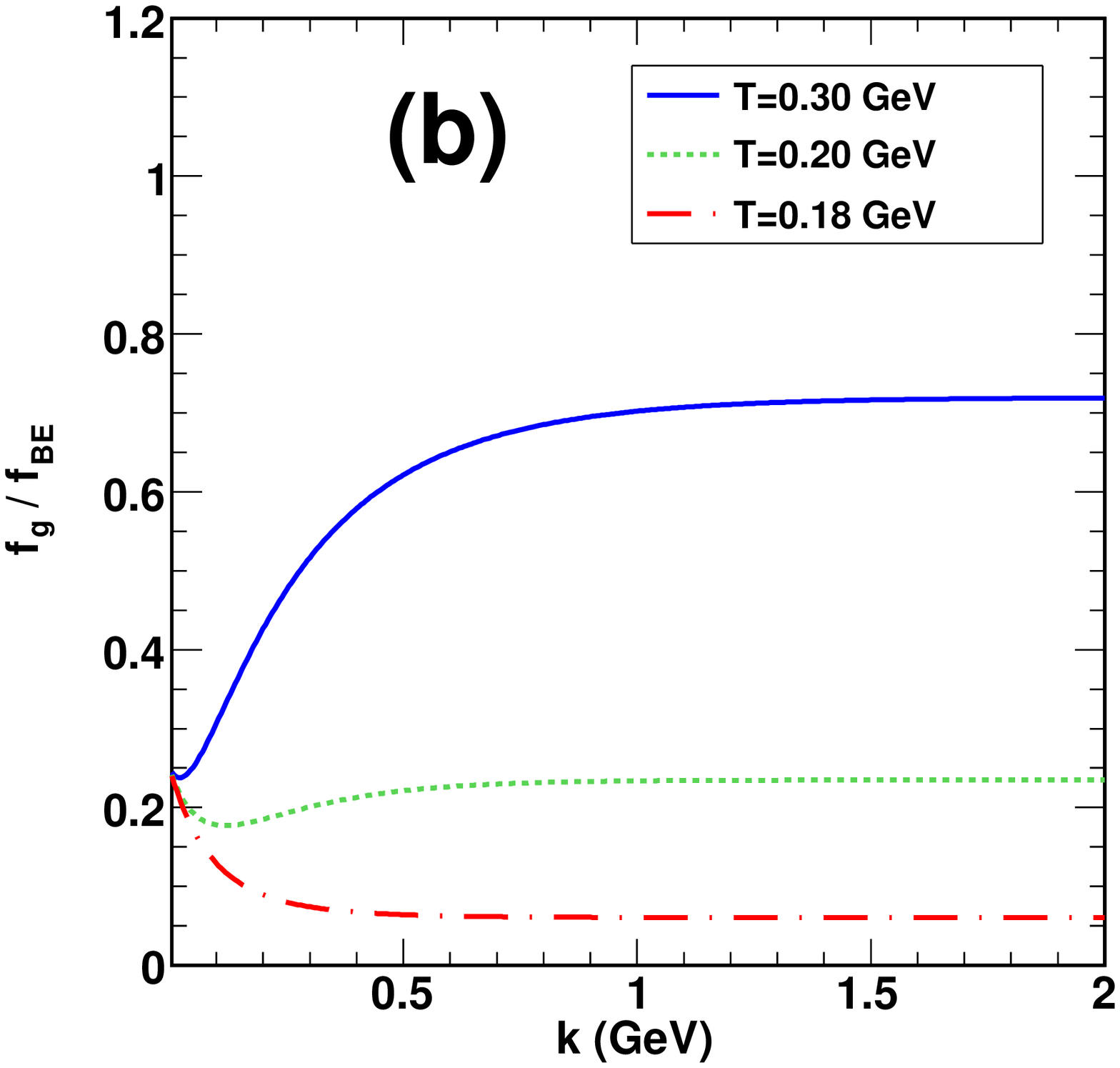,width=7.5cm}} \caption{Ratios of
distribution functions, (a) $f_q(k)/f_{\mathrm{FD}}(k)$, with $u$
and $d$ quark chemical potential $\mu=0.1$~GeV, and (b)
$f_g(k)/f_{\mathrm{BE}}(k)$, at temperatures, $T=0.3$, $0.2$ and
$0.18$~GeV.} \label{fig:distributions}
\end{figure}

\begin{figure}[tb]
\centerline{\epsfig{file=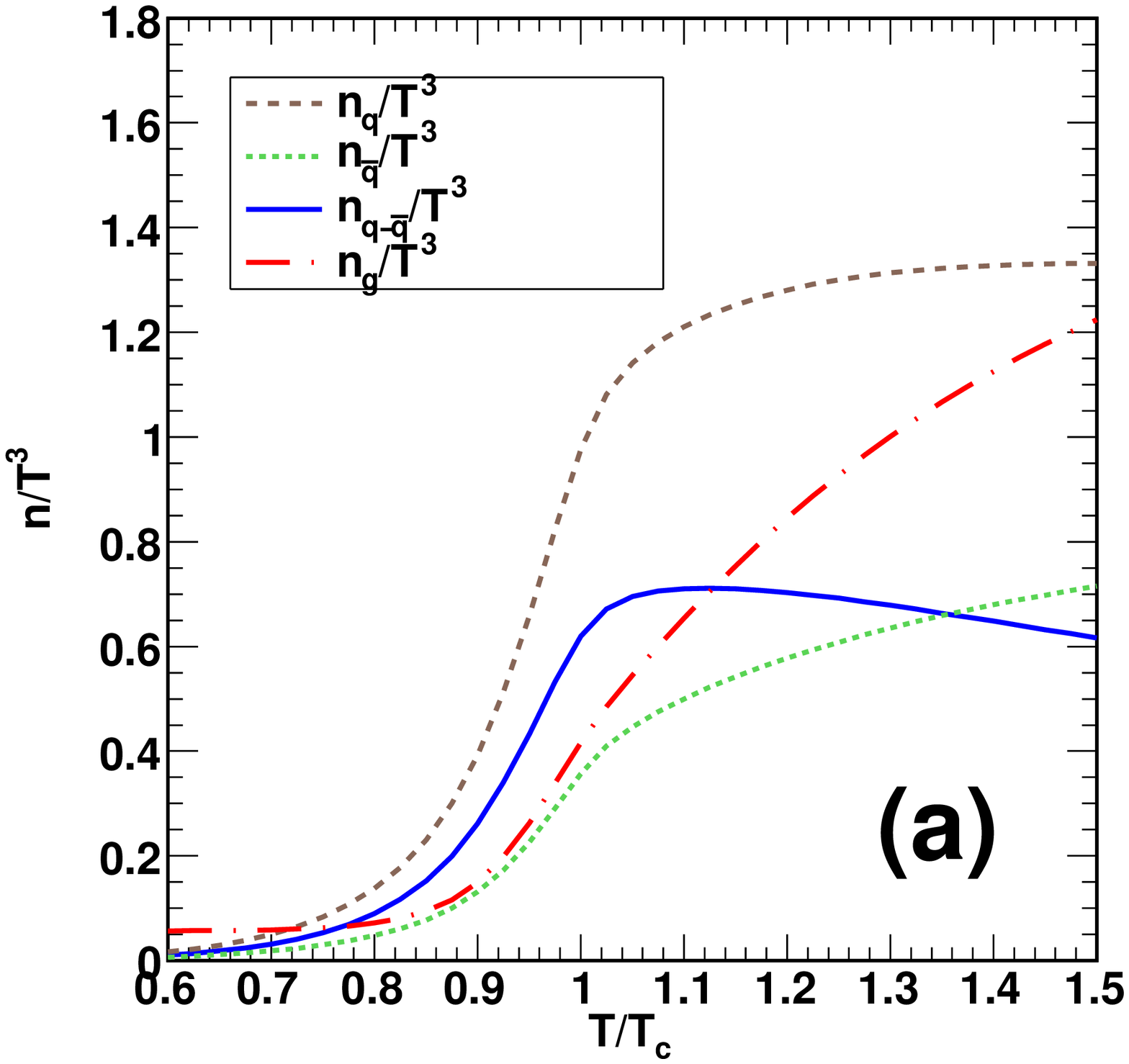,width=7.5cm}
\epsfig{file=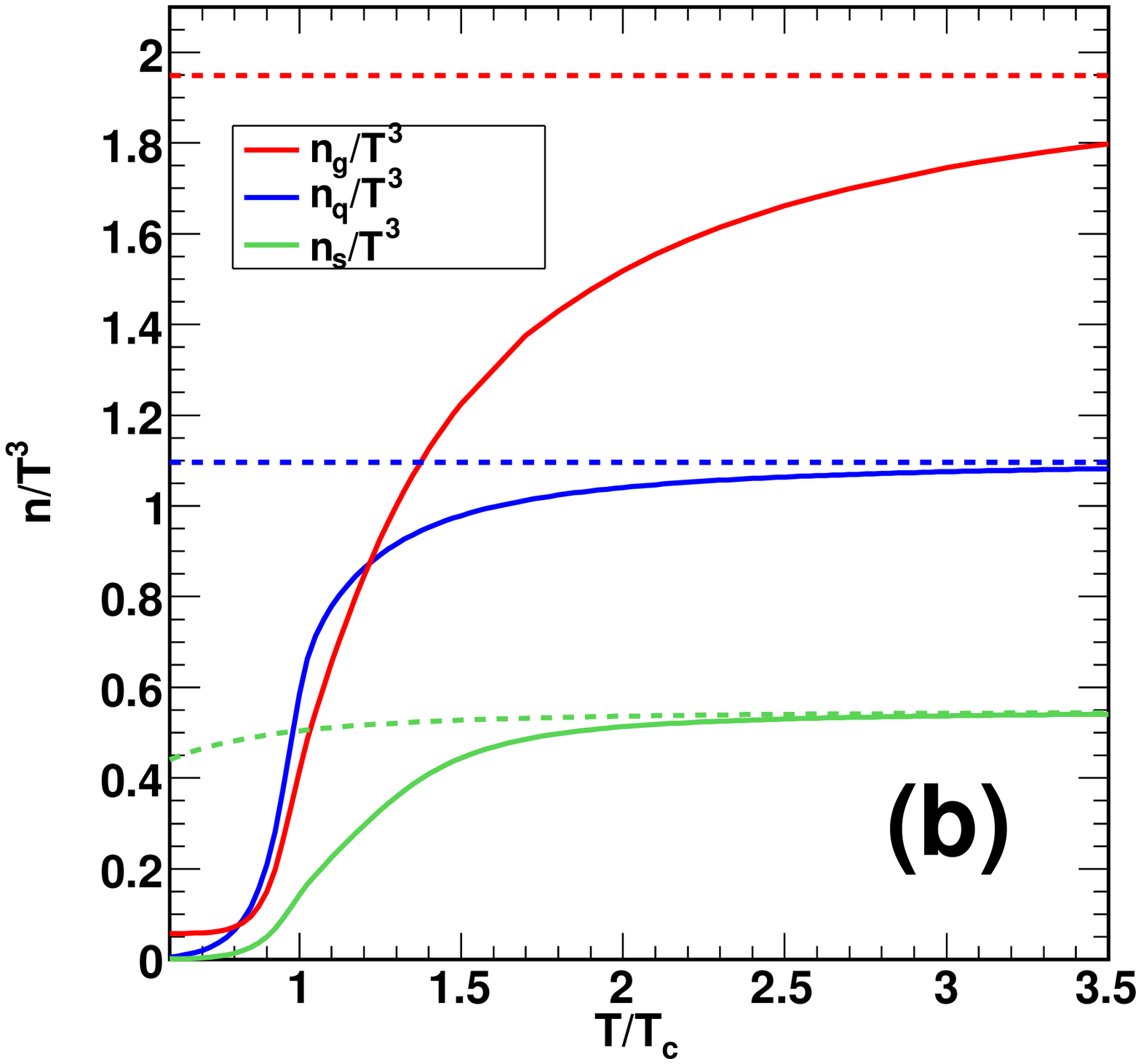,width=7.5cm}} \caption{(a) The temperature
dependence of the (scaled) quark, antiquark, net quark and gluon
number densities, $n_q/T^3$, $n_{\bar{q}}/T^3$, $n_{q-\bar{q}}/T^3$
and $n_g/T^3$, respectively. The $u$ and $d$ quark chemical
potential $\mu=0.1$~GeV. (b) $n_g/T^3$, $n_q/T^3$ and the (scaled)
strange quark number density $n_s/T^3$ as functions of the
temperature, under the assumption of zero quark chemical potentials
for all three flavors, $\mu=\mu_s=0$. In (b), the three dashed
curves represent asymptotic values of $n_g/T^3$, $n_q/T^3$ and
$n_s/T^3$ respectively, for $\langle \ell_3 \rangle = \langle
\bar{\ell}_3 \rangle =\langle \ell_8 \rangle = 1$ and the current
quark masses.} \label{fig:number_density}
\end{figure}

By numerically evaluating $\langle {\tr_A L_8^n } \rangle $ for each
power $n$ and as a function of temperature $T$, the gluon
distribution function can now be obtained from
(\ref{eq:gluon-dist}). The ratios $f_q ( k ) / f_{\mathrm{FD}} ( k
)$ and $f_g \left( k \right) / f_{\mathrm{BE}} \left( k \right)$ are
plotted in figure~\ref{fig:distributions}, where $f_{\mathrm{FD}}$
and $f_{\mathrm{BE}}$ are the free Fermi-Dirac and Bose-Einstein
distributions, respectively, with the constituent quark mass $m_q$
in (\ref{eq:mq}) and $\mu =0.1$~GeV. When $T$ is near $T_c$, gluon
is more strongly suppressed than quarks in spite of the additional
effect of chiral symmetry breaking on the constituent quark mass
$m_q$. Furthermore, we evaluate the quark, antiquark, net quark and
gluon number densities by
\begin{eqnarray}
n_q = n_u+n_d = 4 N_c\int \frac{d^3k}{(2\pi)^3}  f_q(k),
\label{eq:n_q}\\
n_{\bar{q}} = n_{\bar{u}}+n_{\bar{d}} = 4 N_c\int \frac{d^3k}{(2\pi)^3} f_{\bar{q}}(k),
\label{eq:n_q2}\\
n_{q-\bar{q}} = n_q-n_{\bar{q}} = 4 N_c\int \frac{d^3k}{(2\pi)^3} \left[ f_q(k)-f_{\bar{q}}(k) \right],
\label{eq:n_q-q2}\\
n_g = 2 (N_c^2-1)\int \frac{d^3k}{(2\pi)^3}  f_g(k),
\label{eq:n_g}
\end{eqnarray}
where the momentum integration is taken without imposing any cutoff.
The number of flavors in (\ref{eq:n_q})-(\ref{eq:n_q-q2}) is 2
because we are evaluating the quark, antiquark, or net quark number
densities for flavor $u$ and $d$. Figure \ref{fig:number_density}(a)
shows the temperature dependence of the (scaled) quark, antiquark,
net quark and gluon number densities, $n_q/T^3$, $n_{\bar{q}}/T^3$,
$n_{q-\bar{q}}/T^3$ and $n_g/T^3$, respectively. In
figure~\ref{fig:number_density}(a), $n_{q-\bar{q}}/T^3$ possesses
similar features as that in the two-flavor PNJL model
\cite{Ratti:2005jh}. Moreover, $n_g/T^3$ is non-vanishing at low
temperature due to the existence of the color-singlet gluon states.
On the other hand, assuming zero quark chemical potentials for all
three flavors, $\mu=\mu_s=0$, we evaluate $n_q/T^3$, $n_g/T^3$ and
the (scaled) strange quark number density $n_s/T^3$ as functions of
the temperature, as shown in figure~\ref{fig:number_density}(b). We
note that at $T=1.5\,T_c$ the quark densities have reached almost
90\% of their asymptotic values; whereas the gluon density is still
less than 2/3 of its asymptotic value ($n_g/T^3 \approx 1.95$). The
quark and gluon number densities reach 99\% and 92\% of their
asymptotic values respectively at $T = 3.5 T_c$.

\begin{figure}[tb]
\centerline{\epsfig{file=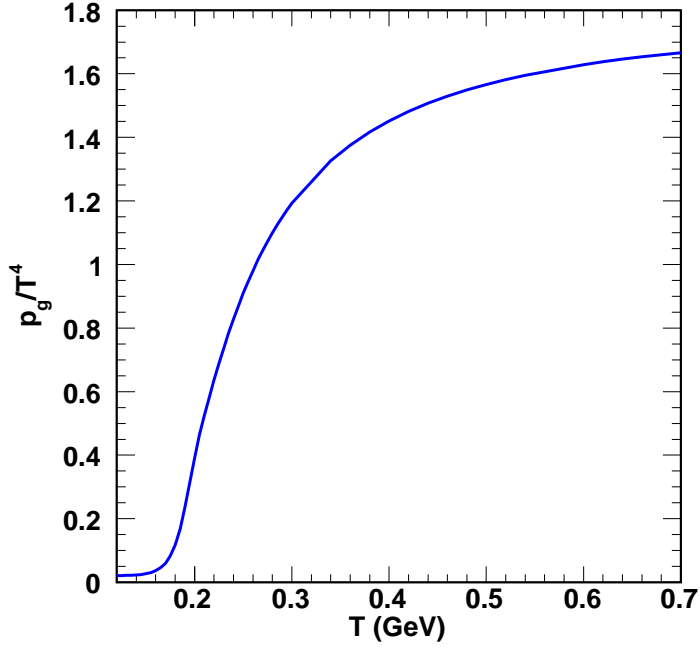,width=10cm}} \caption{The
temperature dependence of the (scaled) thermodynamic pressure,
$p_g/T^4$, contributed by the thermal (transverse) gluons.}
\label{fig:potentials}
\end{figure}

With $f_g(k)$, we can now calculate the contribution of the thermal
(transverse) gluons to the thermodynamic pressure:
\begin{eqnarray}
p_g=\frac{2(N_c^2-1)}{3}\int \frac{d^3k}{(2\pi)^3}  |
\mathrm{\mathbf{k}} | f_g(k),\label{eq:Omega_g_P2}
\end{eqnarray}
\noindent The (scaled) pressure $p_g/T^4$ is plotted as a function
of the temperature in figure~\ref{fig:potentials}. Because the
thermal average of the adjoint Polyakov loop $\langle\ell_8\rangle
\to 1$ for $T \gg T_c$, $p_g/T^4$ approaches the asymptotic behavior
predicted by the Stefan-Boltzmann value, $p_g/T^4
\buildrel{T\to\infty}\over{\longrightarrow} 16\pi^2/90 \approx 1.75$
when $T \gg T_c$. Figure~\ref{fig:potentials} confirms that the
gluon pressure reaches $95\%$ of the Stefan-Boltzmann value at
$T=700$~MeV.

\section{Strange quark pair-production rate}

With the expressions for $f_q ( k )$, $f_{\bar {q}} ( k )$ and $f_g
( k )$, the strange quark pair-production rate for both $q\bar {q}
\to \mbox{s}\bar {s}$ and $gg \to \mbox{s}\bar {s}$ can be derived
in the PNJL model. The strange quark pair-production rate per unit
volume is given by
\begin{eqnarray}
A= \frac{dN}{ dtd^3x} = A_q + A_g,
\end{eqnarray}
\noindent
where
\begin{eqnarray}
A_q &=& \frac{1}{2}\int_{4m_s^2 }^\infty {ds} s\, \sqrt {1 -
\frac{4m_q^2 }{s}} \; \delta \left( {s - \left( {k_1 + k_2 }
\right)^2} \right) \bar {\sigma
}_{q\bar {q} \to s\bar {s}} \left( s \right)
\nonumber\\
&& \times \int \frac{d^3k_1 }{(2\pi )^3  E_q ({\bf k}_1)}
\int {\frac{d^3k_2 }{(2\pi )^3 E_q ({\bf k}_2)}
\left( {2\times 36} \right)} f_q \left( {k_1 } \right)
f_{\bar {q}} \left( {k_2 } \right),
\label{eq:A-q-1}
\\
A_g &=& \frac{1}{2}\int_{4m_s^2 }^\infty {ds} s \, \delta \left(
{s - \left( {k_1 + k_2 } \right)^2} \right) \bar {\sigma }_{gg \to
s\bar {s}} \left( s \right)
\nonumber\\
&& \times \int {\frac{d^3k_1 }{(2\pi )^3\left| {{\mathrm {\bf k}}_1 }
\right|}} \int {\frac{d^3k_2 }{(2\pi )^3\left| {{\mathrm {\bf k}}_2
} \right|}\left( {\frac{1}{2}\times 256} \right)} f_g \left( {k_1 }
\right) f_g \left( {k_2 } \right), \label{eq:A-g-1}
\end{eqnarray}
where $E_q ({\bf k}) = ( \left| {\bf k} \right|^2 + m_q^2 )^{1/2} $.
The cross sections are explicitly given by \noindent
\begin{eqnarray}
\bar {\sigma }_{q\bar {q} \to s\bar {s}} \left( s \right) &=&
\frac{8\pi \alpha _s^2 }{27s^3}\left( {s^2 + 2s\left( {m_q^2 + m_s^2
} \right) + 16m_q^2 m_s^2 } \right)
\nonumber \\
&& \times \left( {1 - \frac{4m_s^2 }{s}}
\right)^{1 / 2} \left( {1 - \frac{4m_q^2 }{s}} \right)^{ - 1 / 2},
\label{eq:sigma-q}
\\
\bar {\sigma }_{gg \to s\bar {s}} \left( s \right) &=& \frac{2\pi
\alpha _s^2 }{3s}\left\{  \left( {1 + \frac{4m_s^2 }{s} +
\frac{m_s^4 }{s^2}} \right) \tanh ^{ - 1}\left[ {\left( {1 -
\frac{4m_s^2 }{s}} \right)^{1 / 2}} \right] \right.
\nonumber\\
&& \left. - \left( {\frac{7}{8} + \frac{31m_s^2 }{8s}} \right)\left( {1
- \frac{4m_s^2 }{s}} \right)^{1 / 2} \right\}. \label{eq:sigma-g}
\end{eqnarray}

\noindent Setting $k_1=|\mathbf{k}_1|$ and $k_2=|\mathbf{k}_2|$, we
can simplify (\ref{eq:A-q-1}) and (\ref{eq:A-g-1}) as follows:
\begin{eqnarray}
A_q &=& \frac{9}{4\pi ^4}\int_{4m_s^2 }^\infty ds \, s \, { \sqrt {1
- \frac{4m_q^2 }{s}} \; \bar {\sigma }_{q\bar {q} \to s\bar {s}}
\left( s \right)} \int_0^\infty dk_1 dk_2 \frac{k_1 k_2} {E_q
(k_1)E_q (k_2) }
\nonumber\\
&& \times \theta \left[ {2\left( {k_1 k_2 + E_q (k_1)E_q (k_2) +
m_q^2 } \right) - s} \right] f_q \left( {k_1 } \right)f_{\bar {q}}
\left( {k_2 } \right), \label{eq:A-q-2}
\\
A_g &=& \frac{4}{\pi ^4}\int_{4m_s^2 }^\infty ds \, s \; \bar
{\sigma }_{gg \to s\bar {s}} \left( s \right)\int_0^\infty dk_1 dk_2
\, \theta \left( {4k_1 k_2 - s} \right)f_g \left( {k_1 } \right)f_g
\left({k_2 } \right). \label{eq:A-g-2}
\end{eqnarray}
By substituting equations~(\ref{eq:quark-dist}),
(\ref{eq:antiquark-dist}), (\ref{eq:gluon-dist}), (\ref{eq:sigma-q})
and (\ref{eq:sigma-g}) into equations~(\ref{eq:A-q-2}) and
(\ref{eq:A-g-2}), the numerical values of the production rates are
obtained as functions of the temperature.

\begin{figure}[tb]
\centerline{\epsfig{file=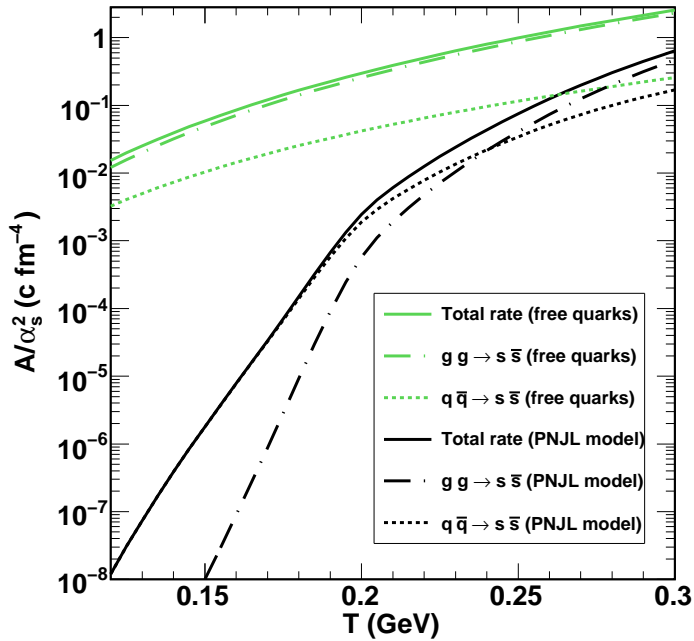,width=10cm}} \caption{Strange
quark pair-production rates divided by $\alpha_s^2$ as functions of
the temperature. The chemical potential for $u$ and $d$ quarks is
$\mu=0.1$~GeV. } \label{fig:rates}
\end{figure}

Figure~\ref{fig:rates} shows the temperature dependence of the
strange quark pair-production rates in the PNJL model, compared with
those obtained for free quarks. One notices a number of qualitative
differences between the rates calculated in the the PNJL model and
those calculated in free perturbation theory. First, the rates are
suppressed for all values of the temperature. This is, in part, due
to the suppression of the thermal quark- and gluon excitations by
the Polyakov loop and, in another part, due to the fact that the
effective strange quark mass remains larger than the current quark
mass even at temperatures moderately above $T_c$, as shown in
figure~\ref{fig:masses}. We also note that the curves for the
Polyakov loop-suppressed gluon induced production rate $A_g$ drops
below the quark induced production rate $A_q$ below $T \approx 240$
MeV, reflecting the stronger suppression of gluons at low
temperature caused by the Casimir scaling of the thermal average of
the adjoint Polyakov loop.

\section{Conclusions}

We have studied the effects of deconfinement and chiral symmetry
breaking on the rates of strange quark pair production in the
framework of the PNJL model. As proposed in
\cite{Rafelski:1982pu,Koch:1986ud}, the strange quark
pair-production rate is enhanced in the deconfined phase
 for the free quarks and the production rate for $gg \to s\bar {s}$ is dominant
at all temperatures. In the PNJL model, the enhanced production of
strange quarks is also obtained, but the production rates for
$q\bar {q} \to s\bar {s}$ and $gg \to s\bar {s}$ cross over at $T_r
\approx 240$MeV. The production rate for $q\bar {q} \to s\bar {s}$
is dominant when $T<T_r$, while that for $gg \to s\bar {s}$ is
dominant when $T>T_r$. Besides, when $T<T_c$, the production rates
for $q\bar {q} \to s\bar {s}$ and $gg \to s\bar {s}$ are both very small
in the PNJL model because quark and gluon quasiparticles are
strongly suppressed below $T_c$. In this temperature region, strangeness
production is dominated by hadronic reactions, which were investigated by
Rehberg {\em et al.} in the NJL model  \cite{Rehberg:1995kh}.

In figure~\ref{fig:rates}, the coupling $\alpha_s$ scales out when
we compare the quark and gluon contributions to the production rate.
We note that, approaching $T_c$, one needs to take into account the
interactions originated from the appearance of collective modes due
to the onset of the spontaneous breaking of chiral symmetry,
rendering our treatment incomplete in the transition region.
However, our goal was to study at what temperature above $T_c$ the
gluonic contribution to the production rate becomes dominant.  We
found that this temperature is around $240$~MeV within the framework
of the PNJL model. Because this threshold is well beyond the
temperature range in which the chiral phase transition occurs, as
can be seen from figure~\ref{fig:order_parameters}, our neglect of
the contribution from collective (hadronic) modes appears justified.

A by-product of our investigation is the demonstration that the
thermal average in (\ref{eq:expectation}) satisfies Casimir scaling
(\ref{eq:Casimir}) of the fundmental and adjoint Polyakov loops.
This gives confidence that the weight function (\ref{eq:weight}) can
be used to obtain the temperature dependence of any quantity that
involves the eigenvalues of the Polyakov loop. For example, the
temperature dependence of $f_g(k)$ requires the evaluation of the
averages $\langle \tr_A L_8^n \rangle$. In turn, $f_g(k)$ makes it
possible to compute the contribution of the thermal (transverse)
gluons to the pressure as a function of the temperature. This makes
it unnecessary to include this contribution explicitly in the
effective potential for the Polyakov loop, as sometimes done in the
literatures \cite{Ratti:2005jh,Roessner:2006xn}.

In our study, the phase transformations of QCD, including
deconfinement and chiral symmetry breaking, are incorporated into
the evaluation of the thermal strange quark pair-production rate.
Using the same techniques, the effects of the Polyakov loop on
other signatures of quark-gluon plasma can be explored in the future.

\section*{Acknowledgements}

This work was supported in part by the U.~S.~Department of Energy
under grant DE-FG02-05ER41367. We thank Kenji Fukushima for several
enlightening discussions about the PNJL model. We thank Inga
Kouznetsova for helpful advice and Johann Rafelski for comments on
the draft of this manuscript. One of us (BM) acknowledges the
hospitality and support of the Yukawa Institute for Theoretical
Physics in Kyoto during the workshop program entitled {\em New
Frontiers in QCD 2008}, which motivated this work.

\appendix
\section{Averaging procedures}\label{sec:appendix}

In this section, we discuss several Polyakov-loop averaging
procedures for the grand canonical thermodynamic potential in the
quark sector and study their implications for the quark and
antiquark distribution functions. This study can be easily extended
to the Polyakov-loop averaging procedures in the gluon thermodynamic
potential, which are not explicitly formulated here. The quark grand
canonical thermodynamic potential per unit volume is defined in
terms of the quark grand partition function,
\begin{eqnarray}
\Omega_q =-\frac{T}{V} \ln \left\langle Z \right \rangle,
\label{eq:full_average}
\end{eqnarray}
where $V$ denotes the volume of the system. In the mean-field
approximation, the quark grand partition function is associated with
a set of quantum numbers, $\alpha=\{\mathbf{k},s,f,c,\pm \}$, where
$\mathbf{k}$, $s$, $f$, $c$ and $\pm$ denotes momentum, spin,
flavor, color and particle/antiparticle quantum number respectively.
The average in (\ref{eq:full_average}) is taken over the eigenvalues
of the Polyakov loop, as shown explicitly in (\ref{eq:expectation}).
Instead of the full average used in (\ref{eq:full_average}), an
approximate averaging method used more frequently in the PNJL model
is the quenched average:
\begin{eqnarray}
\Omega_q \approx - \frac{T}{V} \left\langle \ln Z  \right \rangle =
- \frac{T}{V} \left\langle \ln  \det Z_\alpha  \right \rangle,
\label{eq:quenched_average}
\end{eqnarray}
where $Z_\alpha$ denotes the single-particle partition function
for each quantum number and the determinant runs over all quantum
numbers $\alpha$.
Various further approximations can be applied to
(\ref{eq:quenched_average}), which entail distinct averaging
procedures. In the following texts, we discuss the differences among
several Polyakov-loop averaging procedures for the quark
thermodynamic potential and  clarify their effects on the quark and
antiquark distribution functions.

Define the following subsets of $\alpha$:
$\bar{\alpha}=\{\mathbf{k},s,f,\pm\}$,
$\bar{\alpha}_1=\{\mathbf{k},s,f\}$ and $\bar{\alpha}_2=\{\pm\}$.
Starting from (\ref{eq:quenched_average}), we have the following
(approximate) averaging procedures:
\begin{eqnarray}
\Omega_q &\approx& - \frac{T}{V} \sum_{\bar{\alpha}}  \ln \left
\langle \det_c
Z_{\bar{\alpha},c} \right \rangle, \label{eq:average-1}\\
\bar\Omega_q &\approx& - \frac{T}{V} \sum_{\bar{\alpha}} \left
\langle \ln \det_c Z_{\bar{\alpha},c} \right \rangle,
\label{eq:average-2} \\
\hat\Omega_q &\approx& - \frac{T}{V} \sum_{\bar{\alpha}_1}  \ln
\left\langle \prod_{\bar{\alpha}_2} \det_c
Z_{\bar{\alpha}_1,\bar{\alpha}_2,c} \right \rangle,
\label{eq:average-3}
\end{eqnarray}
where $\det_c$ denotes the color determinant.
Equation~(\ref{eq:average-1}) is the Weiss mean-field approximation,
which is frequently used in the literatures of the PNJL model
\cite{Fukushima:2008wg,Ratti:2005jh}. We note that
(\ref{eq:average-1}) and (\ref{eq:average-2}) take the Polyakov-loop
average for quarks and antiquarks separately. This implies that, in
the limit $\langle \tr_F L_3 \rangle \to 0$, only states with baryon
quantum (quark-triplet) numbers contribute, but not states with
meson quantum numbers (quark-antiquark pairs). We further note that
(\ref{eq:average-1}) and (\ref{eq:average-3}) replace the quenched
average over the Polyakov loop configuration by the unquenched
average. We compare $\Omega_q$ with $\bar\Omega_q$ in
\ref{sec:comparison-1} and $\Omega_q$ with $\hat\Omega_q$ in
\ref{sec:comparison-2}. respectively, by deriving the quark
distribution functions from
equations~(\ref{eq:average-1})-(\ref{eq:average-3}).

\subsection{Validity of the Weiss mean-field approximation} \label{sec:comparison-1}

\begin{figure}[tb]
\centerline{\epsfig{file=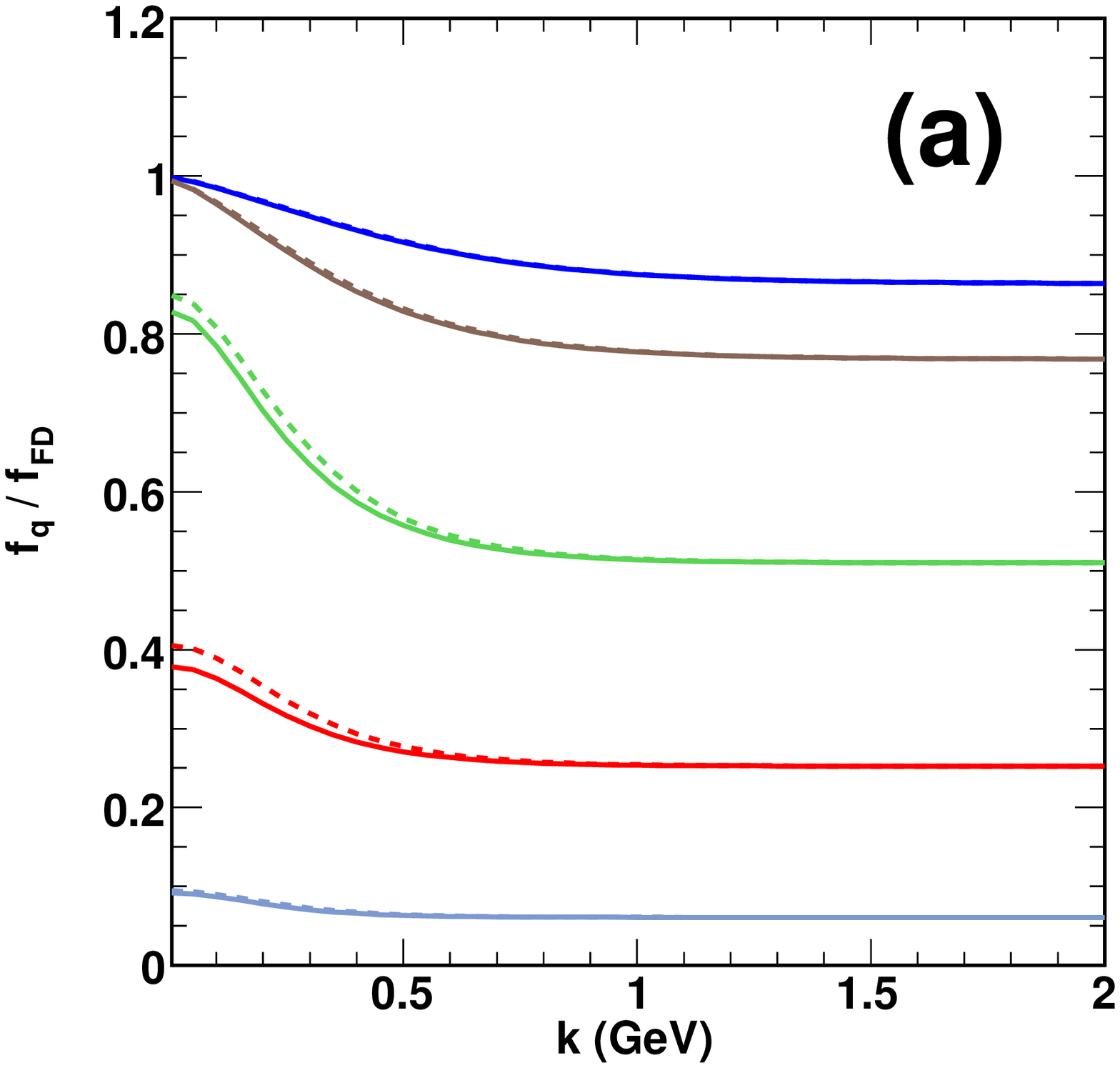,width=7.5cm}
\epsfig{file=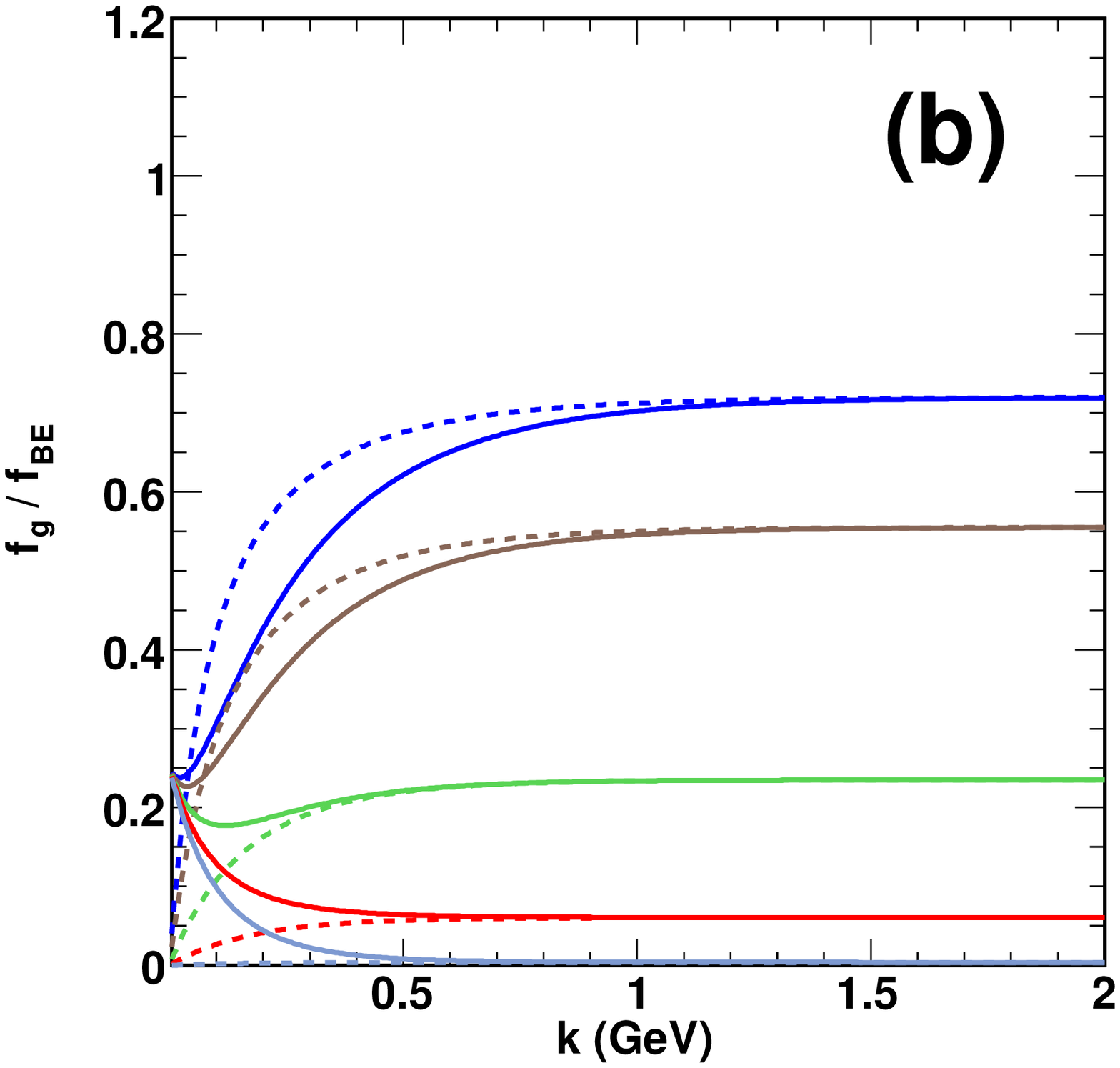,width=7.5cm}} \caption{Validity of the
Weiss approximation for, (a) $f_q(k)/f_{\mathrm{FD}}(k)$, assuming a
zero quark chemical potential, and (b) $f_g(k)/f_{\mathrm{BE}}(k)$.
The solid curves denote the ratios obtained by, (a) $\bar
f_q(k)/f_{\mathrm{FD}}(k)$, where $\bar f_q(k)$ is defined in
(\ref{eq:quark-dist-2}) with $\mu=0$, and (b)
$f_g(k)/f_{\mathrm{BE}}(k)$, where $f_g(k)$ is defined in
(\ref{eq:gluon-dist}). The dashed curves denote the ratios obtained
by, (a) $f_q(k)/f_{\mathrm{FD}}(k)$, where $f_q(k)$ defined in
(\ref{eq:quark-dist}) with $\mu=0$ and $\langle \bar{\ell}_3 \rangle
= \langle \ell_3 \rangle$, and (b) $f_g(k)/f_{\mathrm{BE}}(k)$,
where $f_g(k)$ defined in (\ref{eq:gluon-dist-2}). The curves are
obtained at temperatures, $T=0.3$, $0.25$, $0.2$, $0.18$ and
$0.15$~GeV (top to bottom).} \label{fig:distr-comparison}
\end{figure}

By the averaging procedure in (\ref{eq:average-1}), the quark
thermodynamic potential per unit volume in (\ref{eq:grand-q}) is
simplified to be
\begin{eqnarray}
\Omega _{q} &=& - 2T\sum\limits_{f = u,d,s} \int \frac{d^3k}{(2\pi
)^3} \left\{     \ln \left[ 1 + 3 \langle {\ell _3 } \rangle
\lambda_{+} + 3 \langle {\bar \ell _3 } \rangle \lambda_{+}^2 +
\lambda_{+}^3  \right] \right.
\nonumber\\
&& \left. +   \ln \left[  1 + 3\langle {\bar \ell _3 } \rangle
\lambda_{-} + 3\langle {\ell _3 } \rangle \lambda_{-}^2 +
\lambda_{-}^3 \right]  \right\}
\nonumber\\
&& -6\sum\limits_{f = u,d,s} \int \frac{d^3k}{(2\pi)^3}
E_f(\mathbf{k})  \theta \left(\Lambda ^2 - |\mathrm {\bf k}|^2
\right), \label{eq:grand-q-2}
\end{eqnarray}
where $\lambda_{\pm}=\mathrm{exp}[-((|\mathrm {{\bf k}}|^2 +
m_q^2)^{1 / 2} \mp \mu)/T]$. The quark and antiquark distribution
functions, (\ref{eq:quark-dist}) and (\ref{eq:antiquark-dist}), are
easily obtained from (\ref{eq:grand-q-2}). On the other hand, if we
start from (\ref{eq:grand-q}) and use the averaging procedure
defined in (\ref{eq:average-2}), the quark and antiquark
distribution functions are alternatively obtained:
\begin{eqnarray}
\bar f_{q} \left( k \right) &=& \left \langle \frac{  \ell _3 \lambda_{+}
+ 2 \bar{ \ell} _3   \lambda_{+}^2 + \lambda_{+}^3 }{1 + 3 \ell _3
\lambda_{+} + 3 \bar \ell _3  \lambda_{+}^2 + \lambda_{+}^3} \right
\rangle, \label{eq:quark-dist-2}
\\
\bar f_{\bar {q}} \left( k \right) &=& \left \langle \frac{ \bar \ell _3
\lambda_{-} + 2 \ell _3  \lambda_{-}^2 + \lambda_{-}^3 }{1 + 3\bar
\ell _3 \lambda_{-} + 3 \ell _3 \lambda_{-}^2 + \lambda_{-}^3}
\right \rangle, \label{eq:antiquark-dist-2}
\end{eqnarray}
where $\ell_3$ and $\bar{\ell}_3$ are expressed in
(\ref{eq:ell_3-MFA}). By the definition of the thermal average
(\ref{eq:expectation}) and the weight function (\ref{eq:weight}), we
can evaluate (\ref{eq:quark-dist-2}) and (\ref{eq:antiquark-dist-2})
explicitly. Without losing generality, in this section we assume a
vanishing $u$ and $d$ quark chemical potential, i.e. $\mu=0$, which
implies the simplification $\langle \bar{\ell}_3 \rangle = \langle
\ell_3 \rangle$. Figure \ref{fig:distr-comparison}(a) shows the
comparison of the ratio $f_q(k)/f_{\mathrm{FD}}(k)$ obtained from
(\ref{eq:quark-dist}) with the ratio $\bar
f_q(k)/f_{\mathrm{FD}}(k)$ from (\ref{eq:quark-dist-2}). The figure
shows that the two ratios agree well for all temperatures,
especially in the high momentum region, which is most relevant for
the thermal strange quark pair-production rate.

The gluon distribution function in (\ref{eq:gluon-dist}) is derived
from the gluon thermodynamic potential obtained by the averaging
procedure analogous to (\ref{eq:average-2}). On the other hand, if
using the averaging procedure analogous to (\ref{eq:average-1})
instead, we obtain the gluon distribution function under the Weiss
mean-field approximation, namely
\begin{eqnarray}
f_g \left( k \right) \approx  \frac{ \frac{1}{8} \langle {\tr_A L_8
} \rangle \exp \left( { - {\left| {\mathrm {\bf k}} \right| / T}}
\right)}{1- \frac{1}{8} \langle {\tr_A L_8 } \rangle \exp \left( { -
{\left| {\mathrm {\bf k}} \right| / T}} \right)},
\label{eq:gluon-dist-2}
\end{eqnarray}
\noindent assuming
\begin{eqnarray}
\langle \tr_A L_8^n \rangle \approx 8\left( \frac{1}{8} \langle
 \tr_A L_8 \rangle\right)^n.
\end{eqnarray}
Figure~\ref{fig:distr-comparison}(b) shows the comparison of the
ratio $f_g(k)/f_{\mathrm{BE}}(k)$ obtained from
(\ref{eq:gluon-dist-2}) with that from (\ref{eq:gluon-dist}). They
agree well in the high momentum region but deviate in the low
momentum region.

\begin{figure}[tb]
\centerline{\epsfig{file=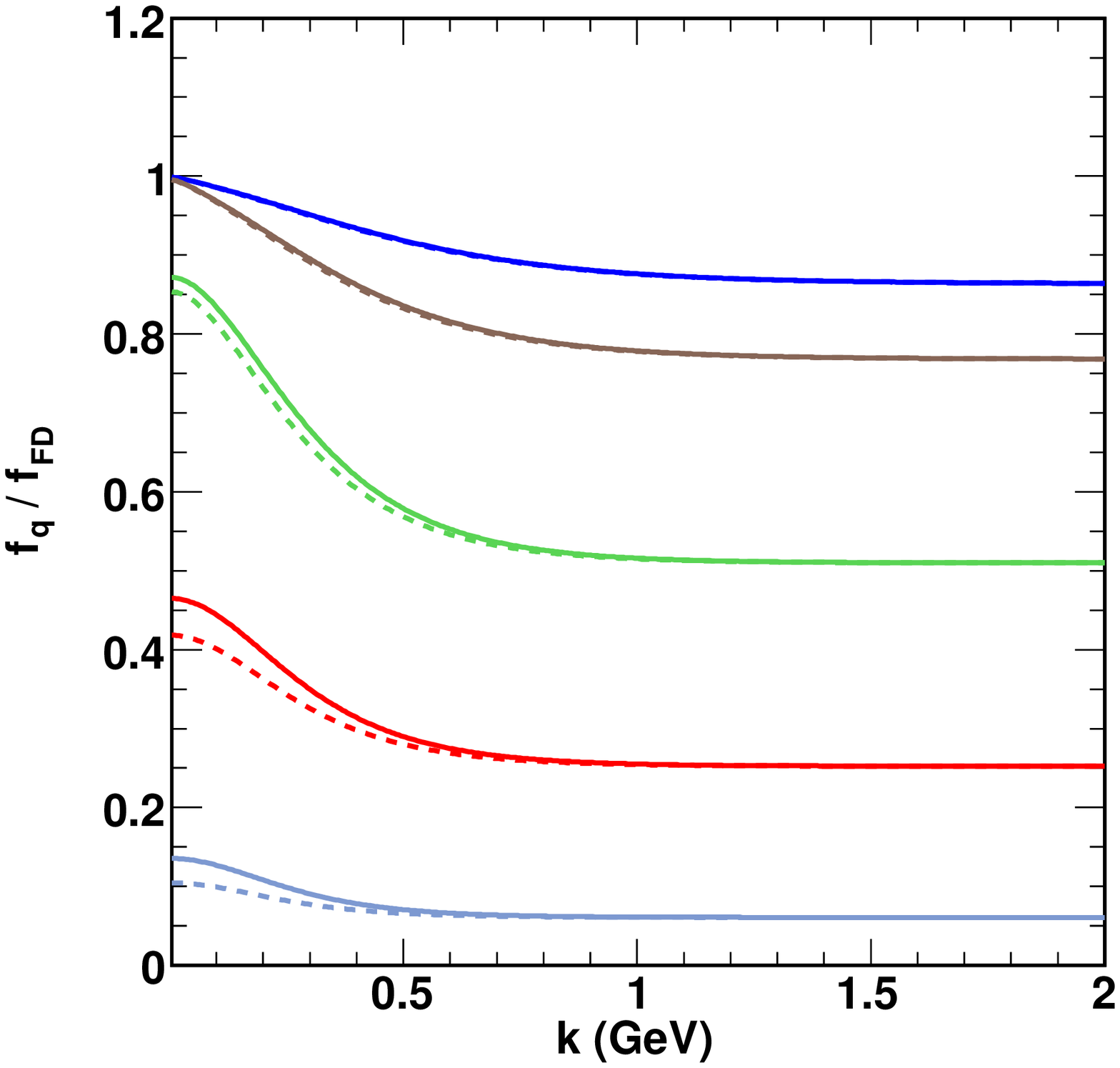,width=10cm}} \caption{The
comparison of the ratio $\hat f_q(k)/f_{\mathrm{FD}}(k)$ obtained
from (\ref{eq:quark-dist-3}), in solid curves, with
$f_q(k)/f_{\mathrm{FD}}(k)$ obtained from (\ref{eq:quark-dist}), in
dashed curves, assuming a quark chemical potential $\mu=0$. The
curves are obtained at temperatures, $T=0.3$, $0.25$, $0.2$, $0.18$
and $0.15$~GeV (top to bottom).} \label{fig:quark-full}
\end{figure}

\subsection{Alternative quark and antiquark distribution functions} \label{sec:comparison-2}

By the Polyakov-loop averaging procedure (\ref{eq:average-3}), we
evaluate alternative quark and antiquark distribution functions,
which, unsurprisingly, incorporate the probabilities of
color-singlet quark-antiquark states.  The quark distribution
function derived from the averaging procedure (\ref{eq:average-3})
is
\begin{eqnarray}
\hat f_{q} \left( k \right) &=& \frac{1}{3} \left[  3
e^{\frac{-6E_q(k)}{T}} + 3 \langle \tr_F L_3 \rangle
e^{\frac{-(5E_q(k)-\mu)}{T}} + 2 \langle \tr_F L_3^\dag \rangle
e^{\frac{-(5E_q(k)+\mu)}{T}} \right. \nonumber \\
&& \left. + 2 \langle \tr_F L_3 \tr_F L_3^\dag \rangle
e^{\frac{-4E_q(k)}{T}}  + 3 \langle \tr_F L_3^\dag \rangle
e^{\frac{-(4E_q(k)-2 \mu)}{T}}  \right. \nonumber\\
&&\left. + \langle \tr_F L_3 \rangle e^{\frac{-(4E_q(k)+2 \mu)}{T}}
+ 2 (2 \langle \tr_F L_3 \rangle +  \langle \tr_F (L_3^\dag)^2
\rangle)
e^{\frac{-(3E_q(k)- \mu)}{T}} \right.   \nonumber\\
&&\left. + (2 \langle \tr_F L_3^\dag \rangle +  \langle \tr_F L_3^2
\rangle) e^{\frac{-(3E_q(k)+ \mu)}{T}} +  \langle \tr_F L_3 \tr_F
L_3^\dag \rangle e^{\frac{-2E_q(k)}{T}}   \right. \nonumber\\
&&\left. + 3 e^{\frac{-3 (E_q(k)-\mu)}{T}}  +2 \langle \tr_F
L_3^\dag \rangle e^{\frac{-2 (E_q(k)-\mu)}{T}} + \langle
\tr_F L_3  \rangle e^{\frac{- (E_q(k)-\mu)}{T}} \right]\nonumber\\
&&\times  \left[ 1+ e^{\frac{-6E_q(k)}{T}} + \langle \tr_F L_3
\rangle e^{\frac{-(5E_q(k)-\mu)}{T}} +\langle \tr_F L_3^\dag
\rangle
e^{\frac{-(5E_q(k)+\mu)}{T}} \right.\nonumber\\
&& \left.  + \langle \tr_F L_3 \tr_F L_3^\dag \rangle
e^{\frac{-4E_q(k)}{T}} + \langle \tr_F L_3^\dag \rangle
e^{\frac{-(4E_q(k)-2 \mu)}{T}}  \right. \nonumber\\
&&\left. + \langle \tr_F L_3 \rangle e^{\frac{-(4E_q(k)+2 \mu)}{T}}
+ (2 \langle \tr_F L_3 \rangle +  \langle
\tr_F (L_3^\dag)^2  \rangle) e^{\frac{-(3E_q(k)- \mu)}{T}} \right.\nonumber\\
&& \left. +(2 \langle \tr_F L_3^\dag  \rangle + \langle \tr_F L_3^2
\rangle)
e^{\frac{-(3E_q(k)+ \mu)}{T}} \right. \nonumber\\
&& \left.  + \langle \tr_F L_3 \tr_F L_3^\dag \rangle
e^{\frac{-2E_q(k)}{T}} +  e^{\frac{-3(E_q(k)-\mu)}{T}} + e^{\frac{-3
(E_q(k) +
\mu)}{T}}     \right.\nonumber\\
&&\left. +  \langle \tr_F L_3^\dag  \rangle e^{\frac{-2
(E_q(k)-\mu)}{T}} + \langle \tr_F L_3  \rangle e^{\frac{-2
(E_q(k)+\mu)}{T}} \right.  \nonumber\\
&& \left. + \langle \tr_F L_3  \rangle e^{\frac{- (E_q(k)-\mu)}{T}}
+ \langle \tr_F L_3^\dag  \rangle e^{\frac{- (E_q(k)+\mu)}{T}}
 \right]^{-1}, \label{eq:quark-dist-3}
\end{eqnarray}
where $E_q(k)=(k^2+m_q^2)^{1/2}$ and $\langle \tr_F L_3 \tr_F
L_3^\dag \rangle = \langle \tr_A L_8 \rangle +1$. Moreover, the
antiquark distribution function $\hat{f}_{\bar{q}}(k)$ can be
obtained from (\ref{eq:quark-dist-3}) by interchanging $L_3$ and
$L_3^\dag$ and replacing $\mu$ by $-\mu$. Equation
(\ref{eq:quark-dist-3}) contains a sum over the probabilities of all
states of $N_1$ quarks and $N_2$ antiquarks, where
$N_1,N_2=\{0,1,2,3\}$. As noted before, (\ref{eq:quark-dist-3})
incorporates the contribution of color-singlet quark-antiquark
states, which are not contained in the expression
(\ref{eq:quark-dist}). Figure \ref{fig:quark-full} shows the
comparison of the ratio $\hat f_q(k)/f_{\mathrm{FD}}(k)$ obtained
from (\ref{eq:quark-dist-3}) with $f_q(k)/f_{\mathrm{FD}}(k)$
obtained from (\ref{eq:quark-dist}), again for $\mu = 0$. The figure
shows that the two ratios agree well for all temperatures,
especially in the high momentum region. We also note that, when $T$
is near $T_c$ or $T<T_c$, $\hat f_q(k)$ in (\ref{eq:quark-dist-3})
is slightly larger than $f_q(k)$ of (\ref{eq:quark-dist}) in the low
momentum region. This difference can be traced back to the
contribution of the color-singlet quark-antiquark states.

In conclusion, because both $\bar f_q(k)$ and $\hat f_q(k)$ are in good
numerical agreement with $f_q(k)$, we are justified to use (\ref{eq:quark-dist})
and (\ref{eq:antiquark-dist}) in the evaluation of the strange quark
pair-production rate, as done in the main part of this article.

\end{document}